\documentclass[jkps,preprint,fleqn,showpacs,showkeys]{revtex4}
\usepackage{graphicx}
\usepackage{amssymb}
\usepackage{amsmath}
\usepackage{bm}
\begin{document}
\setcounter{page}{0}
\title[]{Hubbard physics in the symmetric half-filled periodic Anderson-Hubbard model}
\author{I. Hagym\'asi}
\email{hagymasi.imre@mta.wigner.hu}
\affiliation{Institute for Solid State Physics and Optics, Wigner Research Centre for Physics,
Hungarian Academy of Sciences, Budapest, H-1525 P.O. Box 49, Hungary}
\affiliation{Institute of Physics, E\"otv\"os University, Budapest, P\'azm\'any P\'eter s\'et\'any 1/A, H-1117, Hungary}
\author{K. Itai}
\author{J. S\'olyom}
\affiliation{Institute for Solid State Physics and Optics, Wigner Research Centre for Physics,
Hungarian Academy of Sciences, Budapest, H-1525 P.O. Box 49, Hungary}


\begin{abstract}
Two very different methods -- exact diagonalization on
finite chains and a variational method -- are used to study the possibility of a metal-insulator transition
in the symmetric half-filled periodic Anderson-Hubbard model. With this aim
we calculate the density of doubly occupied $d$ sites ($\nu_d$) as a function of various parameters. 
In the absence of on-site Coulomb interaction ($U_f$) between $f$ electrons, the two methods yield similar results.
The double occupancy of $d$ levels remains always finite just as in the one-dimensional Hubbard model.  
Exact diagonalization on finite chains gives the same result for finite $U_f$, while the Gutzwiller method leads to
a Brinkman-Rice transition at a critical value ($U_d^c$), which depends on $U_f$ and $V$. 
\end{abstract}



\maketitle

\section{INTRODUCTION}

The periodic Anderson-Hubbard model defined by the Hamiltonian
\begin{gather}
\mathcal{H}=\sum_{\boldsymbol{k},\sigma}\varepsilon_d(\boldsymbol{k})
\hat{d}_{\boldsymbol{k},\sigma}^{\dagger}\hat{d}_{\boldsymbol{k},\sigma} 
+ U_{d}\sum_{\boldsymbol{j}} \hat{n}^d_{\boldsymbol{j}\uparrow}\hat{n}^d_{\boldsymbol{j}\downarrow}
  +\varepsilon_f\sum_{\boldsymbol{j},\sigma}\hat{n}^f_{\boldsymbol{j},\sigma}
 +U_f\sum_{\boldsymbol{j}}\hat{n}^f_{\boldsymbol{j}\uparrow}
 \hat{n}^f_{\boldsymbol{j}\downarrow}\nonumber\\
\;\;\;\;\;\;\;\;
-V\sum_{\boldsymbol{j},\sigma}\big(\hat{f}_{\boldsymbol{j},\sigma}^{\dagger}
 \hat{d}_{\boldsymbol{j},\sigma}+\hat{d}_{\boldsymbol{j},\sigma}^{\dagger}
 \hat{f}_{\boldsymbol{j},\sigma}\big)
\end{gather}
is meant to describe the physics of systems in which two types of electrons, one filling a relatively broad conduction band, 
the other a narrow band, are allowed to hybridize. In what follows we call them $d$ and $f$ electrons. The interactions within the bands are denoted by $U_d$ and $U_f$, respectively. In the present work we restrict ourselves to the half-filled paramagnetic case, 
that is, when there are $N_{\uparrow} = N_{\downarrow} = N$ up- and down-spin electrons in 
an arbitrary dimensional lattice with $N$ lattice sites, each of which has one $d$ and one $f$ orbital.
Moreover, we restrict ourselves to the symmetric case, where an equal number of $d$ and $f$ electrons is present on the average. 
This is realized when $\varepsilon_f=(U_d-U_f)/2$, if the energy is measured from the center 
of the $d$ band\cite{Hagymasi:long}.

In our previous study of this model\cite{Hagymasi:long}, in which we used two very different methods: a variational calculation using
the Gutzwiller type wave function and exact diagonalization, 
we were mainly interested in the effect of the on-site interaction $U_d$ 
between conduction electrons on the $f$-electron physics. 
In the present study we examine the effect of $d$-$f$ hybridization ($V$) and of the interaction $U_f$ between $f$ electrons 
on the Hubbard physics, that is on the eventual metal-insulator transition at half filling.

In the Gutzwiller-type treatment of the half-filled Hubbard model, the metal-insulator transition, which is known in this case 
as the Brinkman-Rice transition\cite{BR}, occurs at a finite $U_d$, where the number of doubly occupied $d$ sites becomes zero.
A similar transition was obtained by the Gutzwiller method in the half-filled periodic Anderson-Hubbard model, too\cite{Hagymasi:long},
when the $f$ electrons in the narrow band are strongly correlated. 

In contrast to that, exact diagonalization of the half-filled periodic Anderson-Hubbard model
on finite chains gave a finite number of doubly occupied $d$ sites for any $U_d$, just as in the 
one-dimensional half-filled Hubbard model, where this number is finite for arbitrary $U_d$\cite{Woynarovich}, 
even though the ground state is conducting only for $U_d=0$ and 
insulating for $U_d>0$. 

In this paper we will consider the half-filled symmetric Anderson-Hubbard model in the paramagnetic regime in the full
$U_d\geq 0$, $U_f\geq 0$ sector,
when it is not necessarily in the strongly correlated Kondo region, and will study the possibility of metal-insulator transition.
We should note here that in the present model a ``metallic'' phase is in fact a band insulator with hybridization gap. Therefore, 
we should be speaking about insulator-insulator transition, though its physics is the same as a metal-insulator transition due to 
the interactions between electrons.
We calculate the number of doubly occupied $d$ sites as a function of $U_d$ for various values of $V$ and $U_f$ by both methods 
at the symmetric Anderson point, where the average number of $f$ and $d$ electrons per site (denoted by $n_f$ and $n_d$, respectively)
is exactly 1, and examine the effects of these couplings on the one- and higher dimensional Hubbard physics.

\section{Calculation by exact diagonalization}

First, we perform exact diagonalization of the model on finite chains, where the kinetic energy of conduction electrons moving 
along the chain is described by hopping between nearest-neighbor $d$ orbitals with hopping rate $t$.
  
\begin{figure}[!h]
\centering
\includegraphics[scale=0.65]{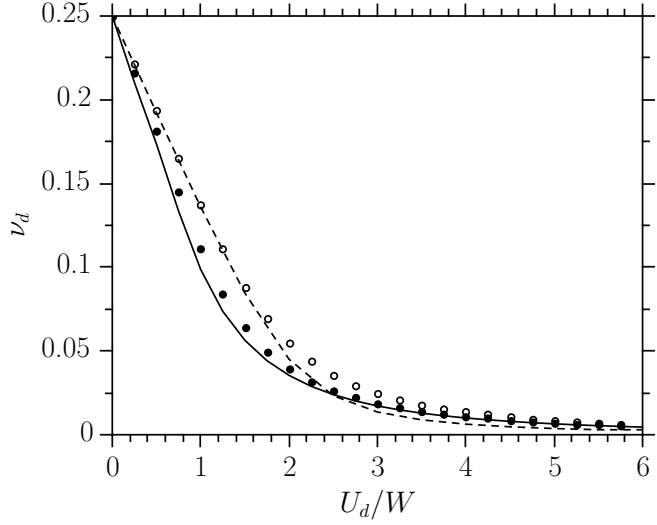}
\caption{\label{nud_Ud_exact:fig} $\nu_d$ as a function of $U_d$ for $V=0.1W$.  The empty and filled circles 
indicate the results of exact diagonalization for $U_f/W=0$ and 5, respectively. 
The dashed line is the result of the Gutzwiller method for $U_f/W=0$. 
The solid line is the exact solution of the one dimensional Hubbard-model.}
\end{figure}
\begin{figure}[!h]
\centering
\includegraphics[scale=0.65]{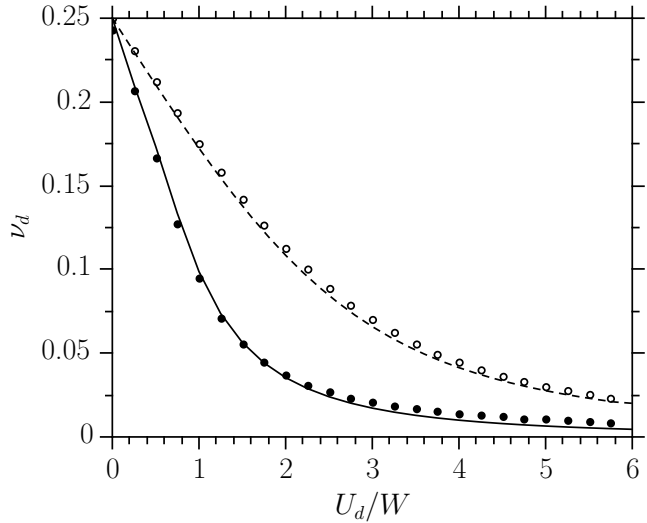}
\caption{\label{nud_Ud_exact_1:fig} The same as Fig. \ref{nud_Ud_exact:fig} except that $V=0.3W$.}
\end{figure}

We consider the periodic Anderson-Hubbard model on a chain consisting of six sites with periodic boundary conditions. 
The results are shown in Figs. \ref{nud_Ud_exact:fig} and \ref{nud_Ud_exact_1:fig}, where the density of doubly occupied $d$ sites, $\nu_d$, is shown as a function of $U_d$ for $V=0.1W$ and $0.3W$, respectively ($W=4t$ is the bandwidth), and for $U_f=0$ and $5W$.
The values calculated with the Bethe Ansatz for the pure Hubbard model are also shown in the figure by a solid line.  

One can see that when the conduction electrons of the $d$ band are hybridized with noninteracting electrons in the $f$ band ($U_f=0$),   the larger the hybridization the more the number of doubly occupied sites. 
The curves are reasonably close to the results obtained by the Gutzwiller method. The agreement gets better for
stronger hybridization, while for weak hybridization it holds for small $U_d$ values only.

The values of $\nu_d$ decrease for finite $U_f$ and get close to those of the pure Hubbard model for large $U_f$. 
This suggests that the $d$-electron subsystem becomes decoupled from the $f$ electrons when the $f$ electrons are strongly correlated.


The results in this section are valid for a chain, for a one-dimensional model. In the next section 
we discuss a variational method, which might be relevant for higher dimensional models.
  
\section{Variational calculation}

We summarize the main steps of the variational calculation following Ref. \cite{Itai:variational}.  
The trial wave function is chosen in the form
\begin{gather} \label{eq:variational_ansatz_Ud}
 |\Psi\rangle=\hat{P}_{\text{G}}^d\hat{P}_{\text{G}}^f\prod_{\boldsymbol{k}}
    \prod_{\sigma}\left[u_{\boldsymbol{k}}
 \hat{f}_{\boldsymbol{k}\sigma}^{\dagger}+v_{\boldsymbol{k}}
 \hat{d}_{\boldsymbol{k}\sigma}^{\dagger}\right]|0\rangle,
\end{gather}
where the Gutzwiller projectors, $\hat{P}^d_{\text{G}}$ and $\hat{P}^f_{\text{G}}$, which contain the variational parameters 
$\eta_d$ and $\eta_f$,  are written as 
\begin{gather}
 \hat{P}_{\text{G}}^d=\prod_{\boldsymbol{j}}\left[1-(1-\eta_d)
    \hat{n}_{j\uparrow}^d\hat{n}_{j\downarrow}^d\right],\\
 \hat{P}_{\text{G}}^f=\prod_{\boldsymbol{j}}\left[1-(1-\eta_f)
    \hat{n}_{j\uparrow}^f\hat{n}_{j\downarrow}^f\right].
\end{gather}
The variational parameters, $\eta_d$ and $\eta_f$, depend on $U_d$ and $U_f$, respectively. Performing 
the optimization with respect to the mixing amplitudes, $u_{\boldsymbol{k}}$ and $v_{\boldsymbol{k}}$, we get 
\begin{gather}
 \mathcal{E}=\frac{1}{N}\sum_{\boldsymbol{k}\in    
     \mathrm{FS}}\left[q_d\varepsilon_d(\boldsymbol{k})+
    \tilde{\varepsilon}_f-\sqrt{\big[q_d\varepsilon_d(\boldsymbol{k})
     -\tilde{\varepsilon}_f\big]^2+4\tilde{V}^2}\right]
      +(\varepsilon_f-\tilde{\varepsilon}_f)n_f+U_d\nu_d+U_f\nu_f
\label{eq:energy}
\end{gather}
for the ground-state energy density, where 
$q_d$ denotes the kinetic energy renormalization factor of $d$ electrons given by 
\begin{eqnarray}
    q_d   =   \frac{1}{\left(1-\frac{n_d}{2}\right)\frac{n_d}{2}} 
    \Bigg[\sqrt{\left(\frac{n_d}{2} -\nu_d\right) \nu_d}
        + \sqrt{\left(\frac{n_d}{2}-\nu_d\right) (1-n_d +\nu_d)}\;\Bigg]^2,
\label{q_d}
\end{eqnarray}
which is formally identical to the expression found in the Hubbard model\cite{Gutzwiller:original}. 
The renormalized hybridization amplitude is now 
$\tilde{V}=V\sqrt{q_dq_f}$; the other notations are the same as 
in our previous paper\cite{Itai:variational}, and the self-consistency condition is given by
\begin{gather}
   n_f = \frac{1}{N}\sum_{\boldsymbol{k}\in \mathrm{FS}} 
   \left[1 + \frac{q_d\varepsilon_d(\boldsymbol{k})-\tilde{\varepsilon}_f}
   {\sqrt{\big[q_d\varepsilon_d(\boldsymbol{k})-\tilde{\varepsilon}_f\big]^2 + 4
   {\tilde V}^2}}\right] = 1.
\label{eq:self-cons2}
\end{gather}
The summation over $\boldsymbol{k}$ can be carried out assuming a constant density of 
states, $\rho(\varepsilon)=1/W$, in the interval $\varepsilon\in[-W/2,W/2]$. 
The values of $\nu_f$, and $\nu_d$ are obtained by 
optimizing the energy density with respect to these parameters numerically. 
For $V\ll W$ and $n_d=n_f=1$ the optimization condition with respect to $\nu_d$ and $\nu_f$ results 
in the following coupled equations:
\begin{gather}
\frac{U_d}{W} - \left[\frac{1}{4}+
2\left(\frac{V}{W}\right)^2\frac{q_f}{q_d}\right]8(1-4\nu_d)=0,\label{eq:nu_d-BR}\\
\frac{U_f}{W} + 2\left(\frac{V}{W}\right)^2\ln{\left[
4\frac{q_f}{q_d}\left(\frac{V}{W}\right)^2\right] }
8(1-4\nu_f) = 0.\label{eq:nu_f}
\end{gather}
When either of $U_d$ or $U_f$ is zero, the equations are decoupled. 
Note that in the absence of hybridization Eq. (\ref{eq:nu_d-BR}) reduces to the result for the ordinary Hubbard model. 

\begin{figure}[!h]
\centering
\includegraphics[scale=0.65]{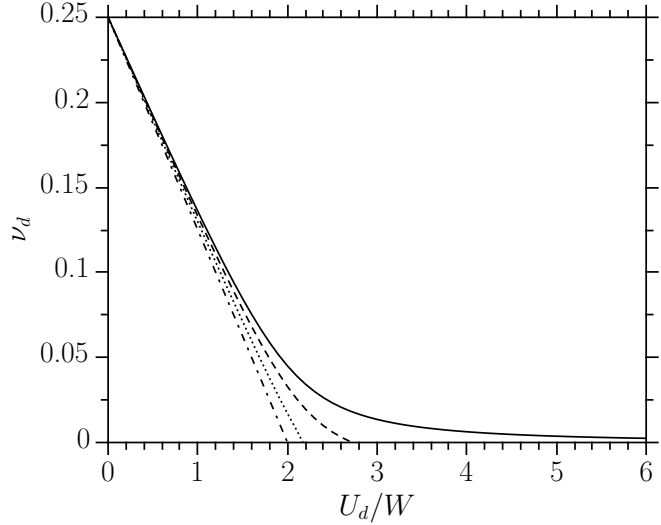}
\caption{\label{nud_Ud:fig} $\nu_d$ as a function of $U_d$. The solid, dashed, dotted and dot-dashed lines correspond to $U_f/W=0,0.3,0.5,5$ respectively. $V=0.1W$ in all cases.}
\end{figure}

\begin{figure}[!h]
\centering
\includegraphics[scale=0.65]{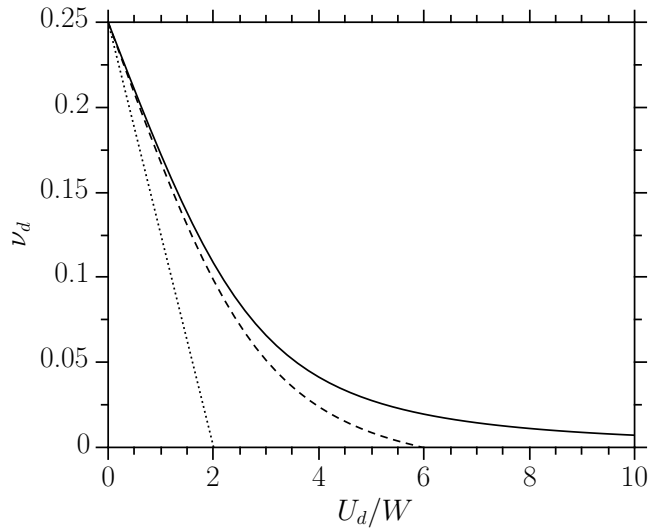}
\caption{\label{nud_Ud1:fig} $\nu_d$ as a function of $U_d$. The solid, dashed and dotted lines correspond to $U_f/W=0,1,10$ respectively. $V=0.3W$ in all cases.}
\end{figure}

We now turn to the discussion of the $d$-electron subsystem. We calculate the density of 
doubly occupied $d$ sites, $\nu_d$, as a function of $U_d$. The results are displayed in Fig. \ref{nud_Ud:fig} for 
several values of $U_f$ for a relatively weak hybridization, $V=0.1W$, and in Fig. \ref{nud_Ud1:fig} for a stronger hybridization.

First, we find that for $U_f=0$, $\nu_d$ never becomes zero, 
just as it was obtained by the exact diagonalization for a finite chain,
that is, the Brinkman-Rice transition does not occur. 
We can calculate the asymptotic behavior of $\nu_d$ for large value of $U_d$ from the following analysis. 
When $U_f=0$, $q_f=1$ irrespective of $U_d$, and Eq. (\ref{eq:nu_d-BR}) can be solved for $\nu_d$. 
For $U_d\gg W$ in leading order we arrive at:
\begin{gather}
 \nu_d\propto \frac{2V^2}{U_d}.
\end{gather}

Second, we find that the Gutzwiller method leads to a Brinkman-Rice transition for any $U_f>0$, 
that is, there exists a finite value, $U_d^c$, 
where $\nu_d$ becomes zero. For $U_d>0$ and $U_f>0$ the optimization conditions, Eqs. (\ref{eq:nu_d-BR}) and (\ref{eq:nu_f}) 
for $\nu_f$ and $\nu_d$ are coupled, and thus both $\nu_d$ and $\nu_f$ decrease when either of the interactions increases.
Therefore, even for very small $U_f$, when $U_d$ is large enough, $\nu_f$ also becomes small, 
and finally both $\nu_d$ and $\nu_f$ simultaneously become zero at $U_d^c$. 
As a matter of fact, $U_d$ and $U_f$ play a rather similar role. 
If we fix $U_d$ at a certain value larger than $2W$, a Brinkman-Rice transition occurs at a certain $U_f^c$.    
The difference in the condition ($U_d>2W$ and $U_f>0$) necessary for occurrence of a transition is due to the different widths of
the $d$ and $f$ bands ($W$ and 0, respectively) in the present model.

Third, when $U_f$ is large enough and $\nu_f$ is exponentially small even for small $U_d$, that is, when the system is the Kondo regime,
the $\nu_d-U_d$ curves become straight lines and coincide to the known behavior of the Hubbard-model. 
In the limit $U_f\gg W$ we obtain:
\begin{gather}
 \nu_d = \frac{1}{4} - \frac{U_d}{8(W+4E_{\text{K}})},\label{nu_d-symm-Ud}
\end{gather}
where
\begin{gather}
E_{\text K} = \frac{W}{2} \exp {\left\{-\frac{U_f}{16V^2/W}\right\}}.
\label{eq:kondo-scale}
\end{gather}

The scenario is the same for weak or strong hybridizations.

\begin{figure}[!h]
\centering
\includegraphics[scale=0.65]{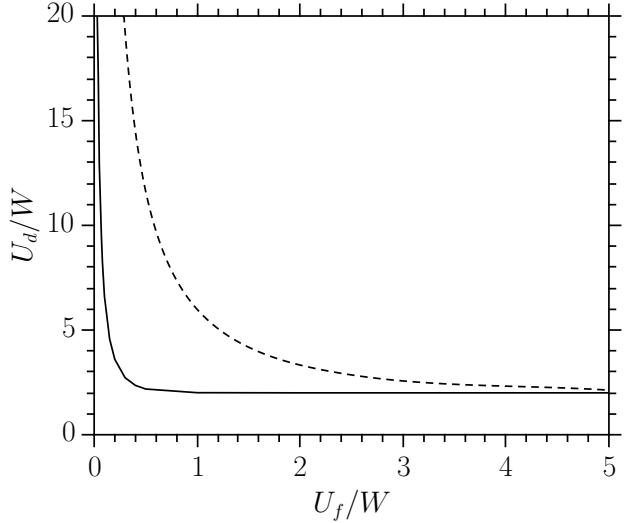}
\caption{\label{phase-dia:fig} The phase boundaries in the $U_d$--$U_f$ plain separating the metallic and insulating regimes. 
The solid and dashed lines correspond to $V=0.1W$ and $0.3W$,
respectively.}
\end{figure}

The results described above indicate that a phase boundary can be defined in the three-dimensional parameter space of $U_d$, $U_f$
and $V$, which separates the region where $\nu_d=\nu_f=0$ from that where both $\nu_d$ and $\nu_f$ are finite. 
The former is a Mott insulator region and the latter is a hybridized band insulator region (for small $U_f$) 
or a Kondo insulator region (for large $U_f$). 
The phase boundaries in the $U_d$--$U_f$ plain are shown in Fig. \ref{phase-dia:fig}.  
(About the boundary between a hybridized band insulator and a Kondo one, see Ref.\cite{Hagymasi:long}.)

\section{Conclusions}

In this paper we discussed the periodic Anderson-Hubbard model focusing our attention 
on the physics of the conduction electron subsystem, 
using two different methods: exact diagonalization on finite chains and a variational method of the Gutzwiller-type.
We studied the effects of the $d$-$f$ hybridization ($V$) and of the on-site interaction between $f$ electrons ($U_f$) 
on the number of doubly occupied $d$ sites, $\nu_d$. 
When $U_f=0$, both methods gave similar results. For larger $U_f$, however, 
the results of exact diagonalization in the one-dimensional model showed that $\nu_d$ approaches that obtained from the Bethe-Ansatz solution of the pure Hubbard model, while the Gutzwiller method indicates 
a Brinkman-Rice-type scenario for a metal-insulator transition.

It is interesting from theoretical point of view that $\nu_d\propto 1/U_d^2$ according to the Bethe-Ansatz solution for $U_d\gg W$, 
while the Gutzwiller method gives a slower asymptotic behavior, $\nu_d\propto 1/U_d$ for $U_f=0$.

\begin{acknowledgments}

This work was supported in part by the Hungarian Research Fund (OTKA) through Grant No. T 68340. One of the authors (I. H.) acknowledges the support of T\'AMOP Grant No. 4.2.2/B-10/1-2010-0030.

\end{acknowledgments}

\end{document}